\documentclass[aps,pre,twocolumn,groupedaddress,showpacs]{revtex4}
\usepackage[]{epsfig}
\newcommand{\la}{\langle}
 
\newcommand{\tr}{t_{\rm rel}}
\newcommand{\upd}{{\mathrm d}}
\newcommand{\eps}{\varepsilon}
\renewcommand{\phi}{\varphi}

\newcommand{\be}{\begin{equation}}
\newcommand{\ee}{\end{equation}}

\begin{document}

\title{A consequence of local equilibration and
heterogeneity in glassy materials}

\author{Ludovic Berthier}
\altaffiliation{Also at:
Laboratoire des Verres, Universit\'e Montpellier II,
34095 Montpellier, France}
\affiliation{Theoretical Physics, 1 Keble Road, Oxford, OX1 3NP, UK}

\date{\today}

\begin{abstract}
The existence of a generalized fluctuation-dissipation theorem 
observed in simulations and experiments performed in various 
glassy materials is related to the concepts of local equilibration
and heterogeneity in space.
Assuming the existence of a dynamic coherence length scale 
up to which the system
is locally equilibrated, we extend previous generalizations of the
FDT relating static to dynamic quantities to the physically relevant 
domain where asymptotic limits of large times and sizes are not reached. 
The formulation relies on a simple scaling argument and has 
thus not the character of a theorem.
Extensive numerical simulations support this proposition.
Our results quite generally apply to systems with slow dynamics,
independently of the space dimensionality, the chosen dynamics,
or the presence of disorder. 
\end{abstract}



\maketitle

\hfill {\it How slow the wind

\hfill How slow the sea} 

\hfill E. Dickinson

\vspace*{0.5cm}

Materials can be far from their equilibrium 
state for two main reasons. First, the relaxation time can be very large
compared to the experimental time scale and the system does not 
reach equilibrium. This is typically the case for
a liquid supercooled through its melting transition below
its glass transition. The glass state is thus 
a non-stationary non-equilibrium state,
where the glass is said to age.
`Glass phases' are extremely commonly encountered in condensed 
matter~\cite{young}. 
Second, when some external drive mechanism is applied 
to the system, it also leaves its equilibrium state. 
Typical situations can be driven interfaces, or soft materials perturbed
by a mechanical flow.
In this paper, we consider these two types of situations
in the specific case of systems with slow dynamics.
We discuss theoretically and numerically
the existence and formulation of a generalization
of the equilibrium fluctuation-dissipation theorem and
its link to equilibrium properties.

Important results for the understanding of
the non-equilibrium dynamics of glassy materials have been obtained in the 
last decade from the asymptotic solution of the dynamics of various 
infinite-range glassy models in the two types of 
situations described above~\cite{review_aging}.
It is established that much information on the dynamics is 
gained by studying two-time correlation and response 
functions, the relationship between them, and the link
between dynamic and static properties.
In the case of an aging dynamics, for instance, 
two-time functions explicitly retain a dependence 
on their two time arguments, and conjugated correlation
and response functions do not satisfy the relation they follow at
equilibrium, the fluctuation-dissipation theorem (FDT).
In the case of a driven dynamics, time translation
invariance might be preserved even in the glassy phase, but
the equilibrium FDT is not satisfied. 

These `mean-field' results have been used as a guide to 
interpret numerical and experimental studies of realistic glassy 
materials~\cite{silvio}. 
However, since no exact solution is known in that case, a naive
mean-field interpretation of experimental results can be misleading.
It is our aim to discuss the interpretation of 
experimental and numerical measurements
of correlation and response functions in terms of static quantities.
For this purpose, we extend previous analysis 
to the physically relevant domain where large times and sizes 
are not accessible. 
Our approach, which relies on the concepts of a spatially local 
equilibration and heterogeneity
of the system, closely follows the line of thought suggested in 
Refs.~\cite{behose,babe,malcom1,malcom2,malcom3,fede,parisi}, 
where the need to take into account spatial aspects of slow dynamics 
was already emphasized in this particular context.

The paper is organized as follows. In Section~\ref{FDTintro}, we 
define the static and dynamics quantities of interest and 
briefly review the known relevant results.  In Section~\ref{FDTscaling}, 
we use a scaling argument to support our generalization of the FDT.  
Numerical simulations of aging and driven dynamics in a spin glass
model in finite spatial dimensions $d=3$ and $d=4$ are reported
in Section~\ref{simu}. Section~\ref{conclu} 
concludes the paper.

\section{Generalized FDT: what do we know?}
\label{FDTintro}

In this section, we introduce the basic concepts and quantities of interest
for this paper. The exact results obtained in infinite-range 
glass models are first presented. We then review the argument 
given to support a possible larger validity of the mean-field results,
and conclude the section with the experimental and numerical 
evidence available. 
The goal of this section is also to recall what is known,
what is only supposed, and most importantly 
to emphasize what is not known.

\subsection{Infinite-range glass models}

The solution of the non-equilibrium dynamics 
of infinite-range glass models consists of the asymptotic analysis
of coupled dynamical equations involving two-time 
correlation functions, $C(t_1,t_2)$, and 
their thermodynamically conjugated response functions, 
$R(t_1,t_2)$~\cite{cuku}. 
Alternatively, integrated responses, or susceptibilities, 
$\chi(t_1,t_2) =  \int_{t_2}^{t_1} \upd t' R(t_1,t')$, can 
be studied, with the advantage that $\chi(t_1,t_2)$ is easier to measure
in a simulation or an experiment.
Adopting notations appropriate to magnetic systems, those functions
read for instance
\begin{eqnarray}
C(t_1,t_2) & = & \frac{1}{N} \sum_{i=1}^N s_i(t_1) s_i(t_2), 
\label{corr}
\\
R(t_1,t_2) & = & \frac{1}{N} \sum_{i=1}^N 
\frac{\partial  s_i(t_1)}{\partial h_i(t_2)} \bigg{|}_{h=0},
\label{resp}
\end{eqnarray}
for a system composed of $N$ spins $s_i \, (i=1, \cdots, N)$. The fields 
$h_i$ are thermodynamically conjugated to the spins.
Since (\ref{corr}) and (\ref{resp}) are self-averaging, no average 
of any kind is required if the thermodynamic limit, $N \to \infty$, 
is taken. 

It turns out that in infinite-range glass models,
a generalized form of the FDT is satisfied in the asymptotic limit
of large times in the aging case, or small driving forces for the 
driven case. 
The limits respectively
read $t_1, t_2 \to \infty$ with $C(t_1,t_2)$ fixed, or 
$\eps \to 0$ with $C(t_1 - t_2)$ fixed, where $\eps$ generically
refers to the amplitude of the driving force (see below for examples).
This asymptotic form of the FDT reads
\be
R(t_1,t_2) = \frac{X[C(t_1,t_2)]}{T} \frac{\partial C(t_1,t_2)}{\partial t_2},
\label{FDR}
\ee
where $T$ is the temperature, and $X(C)$ is called 
the fluctuation-dissipation ratio (FDR). 
The equilibrium 
FDT is recovered when $X(C)=1$.
The finding of a generalized FDT for non-equilibrated systems 
suggests the possibility to develop 
a non-equilibrium extension of standard 
statistical mechanics or thermodynamics. In particular, the 
quantity $T/X$ has been shown to play the role
of a non-equilibrium effective temperature for the slow
modes of the system~\cite{cukupe}, leading to an important body 
of related works~\cite{reviewteff}.

Moreover, in infinite-range glass models 
which statically exhibit a full replica symmetry breaking 
pattern, the new dynamical quantity $X(C)$ was found to be 
directly related
to the static order parameter involved in the replica 
symmetry breaking, the distribution of overlaps
$P(q)$, through the very simple relation~\cite{cuku},
\be 
X(C) = \int_0^C \upd q' P(q').
\label{FDRPQ}
\ee
The `Parisi function' is defined as
\be
P(q) = \lim_{N \to \infty} \left\la \delta \left( 
\frac{1}{N} \sum_{i=1}^N s_i^\alpha s_i^\beta  - q \right) \right\rangle,
\label{PQ}
\ee
where $s_i^\alpha$ denotes the value 
of spin $i$ in the configuration $\alpha$, and $\la \cdots \rangle$ represents
 an average over configurations 
$(\alpha,\beta)$ weighted by their Boltzmann probability.

Equation~(\ref{FDRPQ}) has a remarkable form, since it  relates 
two quantities of completely different origin. The FDR in 
the left hand side of (\ref{FDRPQ}) is measured
in the off-equilibrium dynamics, while the Parisi function 
in the right hand side is computed statically using the equilibrium 
Gibbs measure. 
Note also that the computation of $X(C)$ does not involve replicas.

Finally, we recall the remarkable result that a generalized 
FDT as in Eq.~(\ref{FDR}) 
is asymptotically obeyed in the case of the driven 
dynamics~\cite{cukupe,BBKrheo}, 
with the same value of the FDR, so that Eq.~(\ref{FDRPQ}) remains
unchanged in that case~\cite{BBKultra}.

\subsection{Stochastic stability}
\label{stochastic}

Further developments have taken a more speculative form, in the sense
that no exact result has been obtained for finite dimensional models.
However, it was argued that Eq.~(\ref{FDRPQ}) 
is generally true
for finite dimensional glassy systems~\cite{silvio}. 
We now briefly review the formal derivation of this result to emphasize 
the main hypothesis which are made.
Following Refs.~\cite{cuku},  
the generalized susceptibilities $\chi^r$ are first introduced:
\be
\chi^r = \frac{r!}{N^{r-1}} \sum_{i_1 < \cdots < i_r} 
\frac{\partial  s_{i_1} \cdots s_{i_r}}{\partial
h_{i_1 \cdots i_r}} \bigg{|}_{h=0}  .
\label{sus}
\ee
The `derivation' of Eq.~(\ref{FDRPQ}) 
simply consists of the
comparison between a static and a dynamic computation of 
$\chi^r$~\cite{silvio}. 
A static average gives 
\be
\la \chi^r \rangle_{\rm stat}
= \lim_{N \to \infty}  \lim_{t \to \infty}
\la \chi^r(t) \rangle
= \frac{1}{T} \int_0^1 \upd q P(q) (q^r-1).  
\ee
A dynamical average leads instead to
\be
\la \chi^r \rangle_{\rm dyn} 
= \lim_{t \to \infty} \lim_{N \to \infty}
\la \chi^r (t) \rangle
=  \frac{1}{T} \int_0^1 \upd C \frac{\upd X(C)}{\upd C}
(C^r - 1).
\ee
Equation (\ref{FDRPQ}) directly follows by requiring the equality 
$\la \chi^r \rangle_{\rm stat} =
\la \chi^r \rangle_{\rm dyn}$ for all $r$.

There are, however, several assumptions in this derivation. First, one
assumes that a non-trivial form of the FDT given by Eq.~(\ref{FDR}) 
is valid in the large time limit. Similarly, the static computation is
performed in the thermodynamic limit, $N \to \infty$. 
A third assumption is that static and dynamic calculations
give the same result. This can be justified by a standard nucleation
argument~\cite{silvio}, although exceptions are known~\cite{cuku}.
The last and strongest hypothesis has been called `stochastic 
stability'~\cite{silvio}. 
The name originates from the fact that the susceptibility
(\ref{sus}) probes the behaviour of the system under 
the random perturbation 
\be
H_{\rm perturbation}  = \delta \sum_{i_1 < \cdots < i_r} 
h_{i_1 \cdots i_r} s_{i_1} \cdots s_{i_r},
\ee
in the limit $\delta \to 0$. Here, 
the field $h_{i_1 \cdots i_r}$ is a random Gaussian variable
of mean 0 and variance $r! /(2 N^{r-1})$. 
Stochastic stability implies thus that the limits 
$\delta \to 0$ [or equivalently $h \to 0$ in Eq.~(\ref{sus})],
$N \to \infty$, and $t \to \infty$ can all be inverted.
This last assumption is completely uncontrolled, 
so that Eq.~(\ref{FDRPQ}) is made 
physically plausible by this calculation, 
but has no `theorem' character.
A measurable test of this last hypothesis in a 
realistic system is to check the relation (\ref{FDRPQ}) it implies, 
thus making the argument a bit circular. Interesting
examples and counterexamples were discussed 
in detail in Ref.~\cite{silvio}.

\subsection{Numerical and experimental results}

Since the generalization of the FDT through Eqs.~(\ref{FDR}) and 
(\ref{FDRPQ}) cannot be rigorously proved, numerical simulations
and experiments are necessary in order to make some progress.
The advantage of simulations 
is that both equations can be separately checked, while experiments
can obviously not compare dynamical to static data, since there
are no experimental way to directly measure the Parisi function. 
Its indirect determination through Eq.~(\ref{FDRPQ}) would thus
be a major result, as pointed out in Ref.~\cite{silvio}.
It is the principal aim of this paper to discuss the issue
of what is precisely determined from a dynamical measurement
performed in a physically accessible time window.

Dynamic results have been numerically
obtained in many different glassy models, 
as reviewed in Ref.~\cite{reviewteff}.
More recently, experiments have also been able to 
independently probe two-time susceptibilities
and correlations in a number of glassy 
materials~\cite{manip1,manip2,manip3,manip4}.
The standard way 
to present data and test Eq.~(\ref{FDR}) is to build 
an `FD plot' of the 
susceptibility $\chi(t_1,t_2)$ 
as a function of the correlation function $C(t_1,t_2)$,  
parameterized by the time
difference $t_1 - t_2$, conventionally taken as positive~\cite{cuku}. 
If Eq.~(\ref{FDR}) holds, one gets indeed
\be
\chi(t_1,t_2) = \frac{1}{T} \int_{C(t_1,t_2)}^1 \upd q X(q),
\ee
implying that the $\chi(C)$ relation is independent of time, 
and has a slope related to the FDR, 
\be
\frac{\partial \chi}{\partial C} = -\frac{X(C)}{T}. 
\ee 
We have implicitly assumed that $C(t_1,t_1)=1$ to simplify the
notations.
It is important to notice that neither
simulations nor experiments have ever reported 
an asymptotic FD plot, i.e., a time independent relation
between $\chi$ and $C$.
Quite strikingly then, physically 
accessible time scales are such that Eq.~(\ref{FDR}) is not
valid. As a consequence, it is necessary 
to theoretically investigate preasymptotic behaviours 
where times are large, but finite. 
Equivalently for the driven dynamics, the case of 
small but finite driving forces has to be considered.

Far less works have investigated the validity of Eq.~(\ref{FDRPQ}).
Numerically, it is possible to compute the Parisi function
in an equilibrium simulation, obtaining thus the function $P(q,L)$, where
$L$ is the---necessarily finite---linear 
size of the system. The $L$-dependence is explicitly
kept since it is again important to note that no numerical
work has yet been able to report the asymptotic form of this function,
$P(q,L \to \infty)$, in a glassy system.
Moreover, the system size will play an important role below.

Early studies have then empirically 
compared static and dynamic data through
the comparison of the curves $\chi(C,t_2)$  obtained in the dynamics 
for times $t_2$ 
`as large as possible', to the static curve $S(C,L)$ 
measured statically in a system of size $L$ 
`as large as possible'~\cite{fede2,fede3}. 
We have defined
\be
S(C,L) = \frac{1}{T}
\int_C^1 \upd q \int_0^q \upd q' P(q',L). 
\label{defint}
\ee
With these definitions, Eq.~(\ref{FDRPQ}) can
be synthetically rewritten as
\be
\chi(C) = S(C), 
\ee
where $S(C) = \lim_{L \to \infty} S(C,L)$.

The remarkable coincidence between static and dynamic data 
initially found in
numerical simulations of finite dimensional spin glasses 
in dimensions $d=3$ and $d=4$ was first taken as the sign 
that both quantities had converged to
their asymptotic limits~\cite{fede2}.
Conclusions on the correct description of the low-temperature phase 
in spin glasses phase were then drawn~\cite{fede2}. 

These conclusions were however premature, as strikingly 
demonstrated in recent experiments which have shown that
even in the experimental time window the FD plots still have a clear
time dependence~\cite{manip3}. This again emphasizes the 
necessity of a
careful theoretical study of preasymptotic behaviours
in order to get a correct  
interpretation of experimental data. 
This issue is discussed throughout the rest of the paper.

\section{Generalized FDT with space}
\label{FDTscaling}

\subsection{What about space? A scaling argument}
\label{space}
 
To understand preasymptotic behaviours, a physical picture 
of the slow dynamical processes involved in the system is needed, 
since infinite-range glass models provide us with no 
relevant prediction in that regime. Unfortunately, 
due to their mean-field nature, they do not provide 
us with a clear description of the physics either. One has thus
to resort to more phenomenological descriptions. 

It is now well established, at least in the case of spin glasses, that
slow dynamics is accompanied by the existence of a `coherence
length' scale, $\ell$: the larger the coherence length 
the slower the dynamics.
Here, an analogy with coarsening phenomena in 
non-disordered ferromagnetic systems is quite illuminating. 
When the coarsening dynamics proceeds, larger and larger 
magnetized domains are present, and the dynamics indeed 
becomes slower and slower. 
The extent to which slow dynamics is always accompanied by 
a large coherence length in any
glassy material is an important open
problem in this field. 
In this paper, we adopt the point of view 
that this is indeed the case. See Refs.~\cite{tarjus,zouches} 
for related discussions.

Using this spatial description of slow dynamics, preasymptotic behaviours
were conjectured in Refs.~\cite{behose,babe} 
to obey a generalization of Eq.~(\ref{FDRPQ}).
The physical picture is the following.
At a given wait time, $t_2$, of the aging dynamics,
the system has become locally equilibrated up to
a coherence length scale, $\ell(t_2)$. 
It can thus be seen as a 
heterogeneous mosaic of independent 
and quasi-equilibrated sub-systems of size 
$\ell(t_2)$. Now, a dynamic measurement performed on this mosaic
is in fact probing an ensemble average over 
equilibrated systems of finite size $\ell(t_2)$. 
Translating this idea into an equation, 
one arrives at the conjecture that the 
following generalization of Eq.~(\ref{FDRPQ}) can be 
valid~\cite{behose,babe}:
\be
X ( C,t_2 ) = \int_0^C \upd q' P ( q', L),
\quad L=\ell(t_2),
\label{conj}
\ee 
which implies that non-equilibrium properties of the system at time $t_2$, 
as encoded in the FDR, are related to the equilibrium 
properties of a system of finite size $L=\ell(t_2)$, as 
encoded in the Parisi function.
Note also that static and dynamic 
preasymptotic effects are explicitly present
in Eq.~(\ref{conj}), since both finite times, $t_2$, and sizes, $L$,
are taken into account.

The conjecture (\ref{conj})
was first formulated and tested in Ref.~\cite{behose} in the context
of the non-equilibrium critical dynamics of the 2$d$ XY model, and further
studied in Refs.~\cite{babe,silvio2} 
in the numerical simulation of the Ising spin
glass in dimension $d=2$. In both cases, the observed behaviours
were by construction preasymptotic, since for both models
the true asymptotic behaviour was simple equilibrium with $X(C)=1$ 
and $P(q) = \delta(q)$. However, on numerically 
accessible time scales, a non-trivial behaviour reminiscent
of the one observed, say, in realistic spin glasses was found, 
and the conjectured relation (\ref{conj}) was indeed obeyed to
a very good precision.

It is straightforward to obtain the 
relation corresponding to Eq.~(\ref{conj}) in the case of a driven 
glassy material:
\be
X ( C, \eps ) = \int_0^C \upd q' P ( q', L),
\quad L=\ell(\eps),
\label{conjdriven}
\ee
where the relevant 
control parameter in the dynamics 
is now the amplitude of the driving force, $\eps$.
The advantage of this formulation, as exemplified in the numerical 
simulations below, is that the dynamics is stationary, so that
the coherence length itself does not change during the
dynamical measurement, as is the case in the aging regime. 

\subsection{Single-site quantities}

A different generalization of Eq.~(\ref{FDRPQ}) was recently 
discussed in the literature~\cite{fede,parisi}. 
Inspired by numerical results obtained in a disordered spin
system~\cite{fede}, the argument reviewed in section~\ref{stochastic} 
was reformulated using single-site quantities defined 
for a given sample~\cite{parisi}. 
Therefore, disordered systems only
are concerned by this approach.

Consider the alternative single-site definition of a generalized
susceptibility~\cite{parisi}, 
\be
\chi_i^r = \frac{r!}{M^{r-1}}
 \sum_{a_1 < \cdots < a_r}^M 
\frac{\partial s_i^{a_1} \cdots s_i^{a_r}}{\partial h_{a_1 \cdots a_r}}
\bigg{|}_{h=0},
\ee
where $M$ thermodynamically coupled 
copies of the same system have been introduced, 
so that $s_i^a$ denotes the value of spin $i$ in the copy $a$.
The role of a small coupling 
of amplitude $k$ between the $M$ copies 
is discussed in Ref.~\cite{parisi}. Its presence is necessary to properly
define single-site overlap distributions $P_i(q)$, although
its precise form is inessential. 

If an inversion of the limits $M \to \infty$, $t \to \infty$, $k \to 0$
and $h \to 0$ is again allowed, then a reasoning analogous 
to the one developed in section~\ref{stochastic} leads to the equality
\be 
X_i(C) = \int_0^{C} \upd q' P_i(q'), 
\label{FDRLOCAL}
\ee
between the single-site FDR and local overlap distributions.
The single-site FDR 
is obtained by averaging local correlation, 
$C_i(t_1,t_2) = s_i(t_1) s_i(t_2)$, and response functions,
$R_i(t_1,t_2) = \partial s_i(t_1) / \partial h_i(t_2)$,
over various realizations of the thermal history for a single
realization of the disorder.
We note finally that the derivation of this local FDT 
makes use of the same type of (uncontrolled) 
hypothesis of the global one. 

\subsection{Single-box quantities}

Although Eqs.~(\ref{FDRLOCAL}) and (\ref{conj}) are different, 
they are also consistent, because
Eq.~(\ref{FDRLOCAL}) physically follows
from the concept of a local equilibration in space,
so that dynamical properties at site $i$ can be linked 
to static properties at the same site $i$.

The disadvantage of this single-site formulation
is however evident, since it entirely 
relies on the presence of quenched disorder in the system. 
A slight generalization of Eq.~(\ref{FDRLOCAL}) would be to consider 
single-box quantities instead of single site to get 
similarly:
\be
X_i^v(C) = \int_0^C \upd q' P_i^v(q'),
\label{second}
\ee 
where the box FDR $X_i^v(C)$ is defined from box dynamical functions, 
$C_i^v(t_1,t_2) = v^{-1} \sum _j s_j(t_1) s_j(t_2)$ 
and $\chi_i^v(t_1,t_2) = v^{-1} \sum_j \partial s_j(t_1) /
\partial h_j(t_2)$, and the sums run over a box of 
finite volume $v$ centered around site 
$i$~\cite{malcom1,malcom2,malcom3}. The corresponding 
static box overlap distribution is defined as~\cite{boxdata}:
\be
P_i^v(q) = \lim_{N \to \infty}
\left \langle  \delta \left(
\frac{1}{v} \sum_{j=1}^v s_j^\alpha s_j^\beta - q \right) \right\rangle,
\ee
using the same notations as in (\ref{PQ}). The interest of the
formulation (\ref{second}) is that 
it should become independent of the considered site $i$ 
for a sufficiently large box volume $v$. 
A second interest is that 
it suggests an alternative to Eqs.~(\ref{conj}) and (\ref{conjdriven}), 
namely 
\be 
X(C,t_2) = \int_0^C \upd q' P^v(q), \quad v=\ell^d(t_2),
\label{conjbox}
\ee
for the aging case, and 
\be
X ( C, \eps ) = \int_0^C \upd q' P^v(q'), \quad v=\ell^d(\eps),
\label{conjboxdriven}
\ee
for the driven case.

\section{Simulations of aging and driven dynamics}
\label{simu}

We now turn to a numerical investigation of the two relations 
(\ref{conj}) and (\ref{conjdriven}) discussed in the preceding section.
We first present the model studied and some technical details, before
giving our results both for 
a driven and an aging dynamics.

\subsection{Model and details of the simulations}

In this section, we study numerically the Edwards-Anderson model of a 
spin glass defined by the Hamiltonian~\cite{EA}
\be
H = - \sum_{\langle i,j \rangle}^N J_{ij} s_i s_j,
\label{ham}
\ee
where $s_i \,(i=1 \cdots N)$ are $N=L^d$ Ising spins located at the sites
of a cubic lattice when $d=3$, or a 
hypercubic lattice when $d=4$, of linear size $L$.
The coupling constants $J_{ij}$ are random Gaussian variables
of mean 0 and variance 1 and the sum in (\ref{ham}) runs
over nearest neighbours.
We use a standard Monte Carlo algorithm where the spins are randomly
updated. Times are given in Monte Carlo steps, where one step
represents $N$ attempts to update a spin. 
Sizes are given in units of the lattice spacing.
To study the dynamics of the system, a large system size, $L=40$ ($d=3$)
and $L=24$ ($d=4$), is chosen, so that few realizations of the disorder
are needed, typically 10.
Such large sizes are necessary, as the Fig.~\ref{c4r3d} 
discussed below will clearly demonstrate.
This implies that conclusions drawn 
for systems as small as $L=10$ have to be taken with 
much caution~\cite{star}.

The aging dynamics is simulated by preparing the system at initial
time in a random configuration, 
thus mimicking an infinite temperature state.
The temperature is then changed at time $t=0$ to its final value.
We will present below data for the temperature $T=0.7 T_c$ in both 
$d=3$ and $d=4$, where $T_c=0.95$ and $T_c=1.8$, respectively.
Dynamical measurements are then performed during the 
resulting aging process.

To simulate the driven dynamics, we use the same type of perturbation
as in Ref.~\cite{BBKultra}. On each link $(i,j)$ of the lattice, 
a coupling $\tilde{J}_{ij}$
is added, where $\tilde{J}_{ij}$ is drawn from a Gaussian distribution
of mean 0 and variance $\eps$, which 
defines  then the amplitude of the driving mechanism.
The key information about the $\tilde{J}$'s is that they are chosen
to be antisymmetric, $\tilde{J}_{ij} = - \tilde{J}_{ji}$,
which implies that their effect cannot be incorporated in
a Hamiltonian force of the type (\ref{ham}). 
Hence, the effect of the $\tilde{J}$'s is that of a non-conservative force, 
and the system is thus externally driven. As a result, a 
driven stationary state is reached at sufficiently large times, even
for temperatures below $T_c$: aging is stopped by the driving 
force~\cite{BBKultra}.
In that case, measurements are performed in the stationary
regime, meaning
that a considerable amount of the data corresponding to transient
behaviours has to be discarded.

In both cases, dynamical measurements consist of the measurements
of the global functions (\ref{corr}) and (\ref{resp}). Since these
measurements are by now 
classic, we do not discuss them further here~\cite{reviewteff}.

In order to perform a comparison with static data, 
we shall make use of the 
published equilibrium
data of Ref.~\cite{pqdata3} in $d=3$ and Ref.~\cite{pqdata4} in $d=4$. 
Fortunately, box-overlap data exist in $d=3$ and we shall also 
make use of those~\cite{boxdata}. 

\subsection{Time is length}
\label{prog}

To accurately test the relations (\ref{conj}) and (\ref{conjdriven}),
the following steps must be followed.

\begin{figure}
\begin{center}
\psfig{file=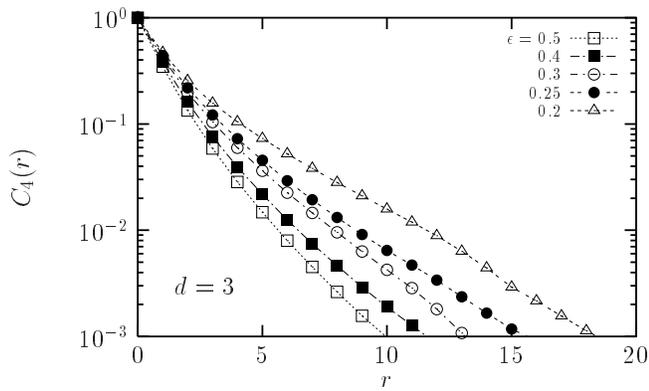,width=8.65cm}
\end{center}
\caption{The correlation
function (\ref{c4}) for $d=3$, $T=0.7 T_c$ and 
various amplitudes $\eps$ of the driving force.}
\label{c4r3d}
\end{figure}

\begin{itemize}

\item Identify and measure 
in the dynamics the coherence length $\ell$.

\item Measure global 
correlation and response functions, from which 
the corresponding FD plot is built.

\item Measure in equilibrium simulations the overlap
distribution functions.

\item Compare static and equilibrium data with carefully chosen
values of the parameters (size, time, driving force), 
as suggested by Eqs.~(\ref{conj}) and (\ref{conjdriven}).

\end{itemize}

The first point, the identification of a coherence length, 
has been discussed for the aging
dynamics of the Ising spin glass in various dimensions in 
Refs.~\cite{heiko,hajime,BB}, 
so that the existence, temperature and wait time behaviour
of this quantity is well-characterized. 

To the best of our knowledge, no such results are available for
the driven dynamics.
We have thus measured in the driven stationary state 
the standard 
four-spin correlation function~\cite{heiko}:
\be
C_4(r) = \frac{1}{N} \sum_{i=1}^N 
\overline{
\langle s_i^a s_{i+r}^a s_i^b s_{i+r}^b \rangle},
\label{c4}
\ee
where the overline means an average over the disorder, 
and $(a,b)$ are two independent copies of the system.
Some corresponding curves are shown in Fig.~\ref{c4r3d}, for 
$d=3$. Very similar results are also obtained for $d=4$.
The features observed in this figure are expected. The amplitude
of the driving force is known to control the relaxation
time, $\tr = \tr(\eps)$, of the system, the smaller 
$\eps$ the larger $\tr$~\cite{BBKultra}. 
From Fig.~\ref{c4r3d}, we recognize 
that a slower dynamics also implies
the existence of a larger coherence length, $\ell(\eps)$, 
as revealed by a slower spatial decay of $C_4(r)$. 

\begin{figure}
\begin{center}
\psfig{file=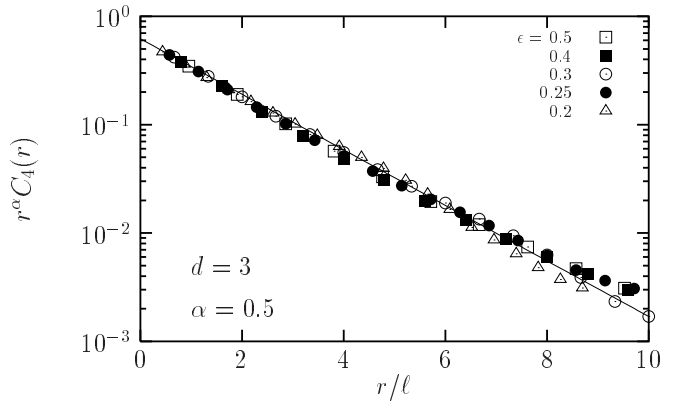,width=8.65cm}
\end{center}
\caption{The data of Fig.~\ref{c4r3d} rescaled 
according to Eq.~(\ref{scal}) with $\alpha = 0.5$, $T=0.7 T_c$, 
$d=3$. The straight line indicates a simple
exponential behaviour of the scaling function 
${\cal C}(x) = \exp(-x)$.}
\label{c4r3d2}
\end{figure}

A non-ambiguous definition of the coherence length
is obtained, as in the aging
regime, through the study of the scaling properties of $C_4(r)$.
For the aging, it is known that the correlation (\ref{c4}) is 
well described by the scaling form~\cite{BB,enzo}
\be
C_4(r) \approx \frac{1}{r^{\alpha(T)}} {\cal C} \left(
\frac{r}{\ell} \right),
\label{scal}
\ee
which can be seen as the definition of
the coherence length $\ell$. We find that the same
scaling behaviour (\ref{scal}) is obeyed in the driven dynamics as well, 
with the same value of the temperature-dependent 
exponent $\alpha(T)$ as in the aging regime. As an example, we report
the scaled data for $T=0.7 T_c$ and $d=3$ in Fig.~\ref{c4r3d2}.
We find that the scaling function ${\cal C}(x)$ is well 
described by a simple exponential form, 
${\cal C}(x) = \exp(-x)$. This contrasts with the `compressed' 
exponential form reported in the aging regime~\cite{BB}.
We have no explanation for this difference, which is
however inessential. 

The results of Figs.~\ref{c4r3d} and \ref{c4r3d2} allow us
to obtain the relationship between the coherence length $\ell$ and
the amplitude of the driving force, $\eps$. From the
time dependence of the correlator (\ref{corr}), it is also possible
to extract a relaxation time, $\tr(\eps)$. It is thus natural
to eliminate $\eps$ to obtain the relationship between
time and length. Not surprisingly, we find
that the relation $\ell(\tr)$ in the driven regime, 
is well compatible with the relation $\ell(t_2)$ measured in the aging
regime at the same temperature.

Although this coincidence might seem anecdotic in this context, we
emphasize its physical importance. This suggests indeed that whatever
the control parameter for the dynamics is, 
waiting time or driving force, the dynamics
is such that the same 
dynamic scaling $\ell(t)$ is obeyed: `Time is length'.

\subsection{Generalized FDT for the driven dynamics}

Following the program presented
in section~\ref{prog}, we compute
in the driven dynamics correlations and susceptibilities, and build 
the corresponding FD plots. Collecting the static data of 
Refs.~\cite{boxdata,pqdata3,pqdata4}, 
we are in a position to start and investigate the validity of 
the relations (\ref{conj}), (\ref{conjdriven}), 
(\ref{conjbox}) and (\ref{conjboxdriven}).

\begin{figure}
\begin{center}
\psfig{file=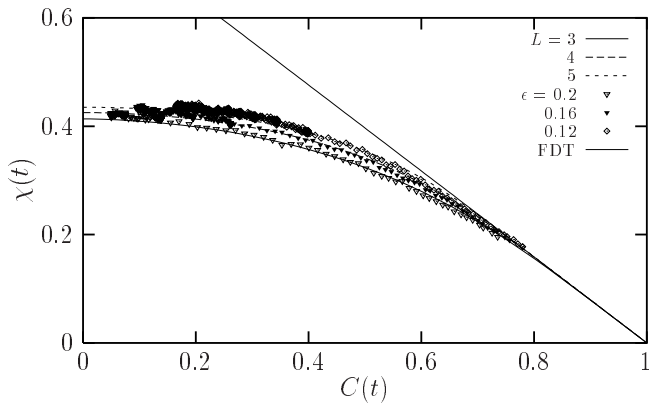,width=8.65cm}
\end{center}
\caption{Test of the static - dynamic relationship (\ref{conjdriven})
for the driven dynamics for $d=4$ and $T=0.7 T_c$.
The points are dynamic FD plots, the lines are the double integral of 
the Parisi function measured in equilibrium simulations.}
\label{param4d}
\end{figure}

We present in Fig.~\ref{param4d} the results obtained 
in $d=4$. In this figure, there are three different FD
plots, represented with points, and 
corresponding to three different values of the driving force, and thus
to three different coherence lengths $\ell(\eps)$. 
The lines correspond to the double integral $S(C,L)$ 
defined in Eq.~(\ref{defint}) of the Parisi function $P(q,L)$
for three different system sizes.
The agreement between the two sets of curves is extremely good. 
That this is not a mere coincidence is
supported by three important facts.

\begin{itemize}

\item
The two types of measurements are performed entirely
independently, in completely different physical situations; 

\item The agreement between the two sets of curves is very good 
on the whole range of the curve, not on a single point. If one 
remembers that each curve is built from a rather complex
manipulation of various quantities without any fitting
procedure or free parameter,
the agreement becomes much 
more impressive;

\item The coherence lengths measured in the dynamics
are $\ell(0.2)=1.55$, $\ell(0.16)=2.0$, $\ell(0.12)=2.5$. 
They are thus in ratio
1:1.29:1.61. The corresponding static length scales are in ratio
1:1.33:1.66. The agreement with the prediction (\ref{conjdriven})
that both lengths should scale similarly 
is thus clearly excellent.

\end{itemize}

\subsection{Another observable: The link overlap}

\begin{figure}
\begin{center}
\psfig{file=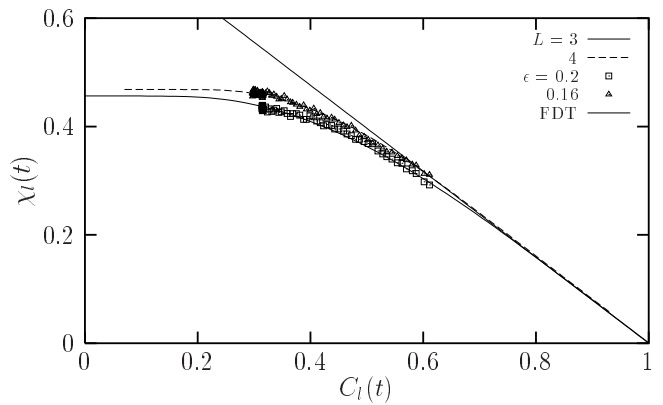,width=8.65cm}
\end{center}
\caption{
Test of the static - dynamic relationship (\ref{conjdriven})
for the driven dynamics for $d=4$ and $T=0.7 T_c$ 
for the link observables.
The points are dynamic FD plots, the lines are the double integral of 
the link overlap distribution function measured in equilibrium simulations.
Note the coincidence between the values of $\eps$ 
and $L$ with the ones reported in Fig.~\ref{param4d}.}
\label{param4dl}
\end{figure}

To test further the hypothesis of a strong link between 
off-equilibrium properties and static ones resulting
from the local equilibration of the material, we focus
in this section on a second pair of dynamical observables.
Physically, this is motivated by the picture
given in section \ref{space}, where an 
off-equilibrium  system was described
as a mosaic of equilibrated sub-systems of size given by the coherence length.
A logical consequence is that the link between static and dynamic 
quantities should exist for all physical observables. 
This is equivalent to the prediction 
and test of the unicity of the effective 
temperature in structural glasses~\cite{cukupe,sollich,shear,mayer}.

To check this hypothesis, we have investigated the following 
quantities:
\begin{eqnarray}
C_l(t_1,t_2) & = & \frac{1}{2 N d} \sum_{i=1}^N {\sum_{j=1}^{2d}} ' 
s_i(t_1) s_j(t_1) s_i(t_2) s_j(t_2) , \\
R_l(t_1,t_2) & = & \frac{1}{2 N d} \sum_{i=1}^N {\sum_{j=1}^{2d}} ' 
\frac{\partial s_i(t_1) s_j(t_1)}{\partial h_{ij} (t_2)} \bigg{|}_{h=0},
\end{eqnarray}
where the prime in the second sum 
indicates that it is restricted to the nearest neighbours
of the site $i$. The field $h_{ij}$ is thermodynamically conjugated 
to the observable $s_i s_j$, and can thus be seen 
as a random perturbation of the coupling constants.
From these two dynamical functions, a new FD plot can be built.

Correspondingly, the static quantity to use is the link-overlap 
distribution function, $P(q_l)$, defined 
by
\be
P(q_l) =
\lim_{N \to \infty} \left\la \delta \left( 
\frac{1}{2Nd} \sum_{i=1}^N 
{\sum_{j=1}^{2d}} ' 
s_i^\alpha s_j^\alpha
s_i^\beta s_j^\beta
 - q_l \right) \right\rangle.
\ee
We use here the static data published in Ref.~\cite{pqdata4}  

The comparison between the new sets of static and dynamic quantities is
reported in Fig.~\ref{param4dl}, for the same parameters as in 
Fig.~\ref{param4d}.
It should not come as a surprise that the range of correlators covered
by Fig.~\ref{param4dl} is smaller than the one covered in Fig.~\ref{param4d}.
The long-time limit of the correlator is indeed non-zero in that case. 

It is again clear that the agreement between static and dynamic data
is very good. The point to emphasize is that for a given amplitude of the 
driving force, the same size $L$ as in Fig.~\ref{param4d} is used 
to compute static data, in agreement with the physical picture
described above. 

These results are a striking confirmation 
of the local equilibration hypothesis formulated in this paper.

\subsection{Using the box overlap distributions}

Having established numerically the validity of the relationship 
between static and dynamic properties in $d=4$, we now turn 
to $d=3$. This will allow us to make use of the box-overlap 
distribution functions published in Ref.~\cite{boxdata}.
To compute this function, the linear 
size of the total system  was fixed to $L=12$,
and the linear size of the boxes, $B$, was varied~\cite{boxdata}.  

\begin{figure}
\begin{center}
\psfig{file=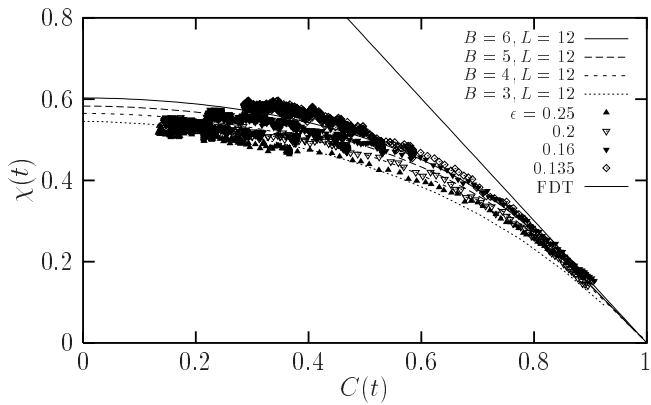,width=8.65cm}
\end{center}
\caption{
Test of the static - dynamic relationship (\ref{conjboxdriven})
for the driven dynamics for $d=3$ and $T=0.7 T_c$.
The points are dynamic FD plots, the lines are the double integral of 
the box-overlap distribution function 
function measured in equilibrium simulations, $B$ refers to
the linear size of the box, $L$ to the linear size
of the total system.}
\label{param3dbox}
\end{figure}

Our numerical results are reproduced in Fig.~\ref{param3dbox}, where
the FD plots built from the
spin-spin dynamical functions (\ref{corr}) and (\ref{resp})
for various amplitudes of the driving force are compared
to the static data obtained from the box-overlap distributions.

As for $d=4$, we find a very good agreement between 
static and dynamic data. 
The coherence length scales are in ratio 1:1.35:1.59:1.88, while
the size of the boxes are in ratio 1:1.33:1.66:2, the coincidence
being again satisfying. 

We conclude that numerical simulations in $d=3$ and $d=4$
of the driven dynamics of the Ising spin glass 
nicely support the theoretical propositions
(\ref{conjdriven}) and (\ref{conjboxdriven}).

\subsection{Generalized FDT in aging dynamics}

We conclude the report of our numerical results with a brief mention
of the
results we have
obtained in the aging regime for $d=3,4$.
Similar results have already been published, as mentioned above, 
but the crucial coincidence between time 
(in the dynamics) and size (in the statics)
was not noted, although it was supported by the data~\cite{fede2,enzo}. 
We have also discussed above the results obtained for $d=2$, 
when no spin glass phase exists~\cite{babe,silvio2}.

\begin{figure}[t]
\begin{center}
\psfig{file=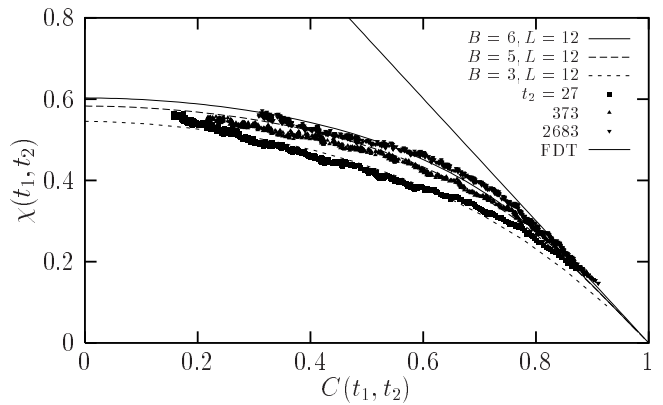,width=8.65cm}
\end{center}
\caption{Test of the static - dynamic relationship (\ref{conjbox}) 
for the aging dynamics, for $d=3$ and $T=0.7 T_c$. The points are
dynamic FD plots, the lines are the double integral 
of the box overlap distributions measured at equilibrium.
The discrepancies for the 
short waiting time $t_2=27$ are discussed in the text.}
\label{aging3d}
\end{figure}

We give an example of our own data in the aging regime 
in $d=3$ in Fig.~\ref{aging3d}.
There are four important 
points we want to make on these data.

First, we make use in this figure of the box-overlap distribution 
function, which was not the case in previous works. 
The total overlap distribution functions give however 
results which are similarly good.

Second, the dynamic data can be described using the same equilibrium
quantity $P(q,L)$ as for the driven dynamics. This is a clear support
of the analytical result that driven and aging dynamics 
should give rise to similar deviations from the 
FDT~\cite{cukupe,BBKrheo,BBKultra}.

Third, we have intentionally included data for a 
very short wait time, $t_2 =27$, for which the agreement with
Eq.~(\ref{conjbox}) is not quite as good, 
to underline the fact that during the
dynamical measurements the coherence length itself evolves with time.
This was obviously not the case when the driven dynamics was 
considered. For this short wait time, for instance, the final
time is $t_1 = 10^5$, so that we do not expect the FD plot
to follow the line built from static data obtained 
at a given box size corresponding to the earlier time 
$L=\ell(t_2)$.
Indeed, a clear deviation at large times 
(corresponding to low values
of the correlator) is observed. This was anticipated 
and noted in Refs.~\cite{behose,babe}, and was another 
motivation to focus also
on the driven dynamics in this paper. 

Fourth, we have discussed above 
the link between the coherence length $\ell$
and the system size or box size in terms of relative ratio.
It is clear that an absolute comparison is not possible, since 
the coherence length has no `absolute value', contrary to the
box-size univocally defined in the static computation.
However, 
the `large' numerical factor between $L$ and $\ell$ reported 
in Ref.~\cite{fede2} (about 4 in $d=3$) was an argument used against the 
theoretical proposition (\ref{conj}) formulated in this 
and previous works~\cite{mezard}.
It can be seen from Fig.~\ref{aging3d} that the factor 
between $\ell$ and $B$ is less than 2 when the box overlap distribution
is used, thus clearly weakening this criticism.

\section{Conclusion}
\label{conclu}

In this paper, we have given a scaling argument to support
the proposition of Refs.~\cite{behose,babe}
to generalize the relationship between
the off-equilibrium FDR and the equilibrium overlap distribution
to the physically relevant domain of finite times and sizes.
Inspired by
previous approaches to the asymptotic relation~\cite{silvio,parisi}, we
proposed the relations (\ref{conj}) and (\ref{conjdriven})
between the FDR computed at any moment of the dynamics, and 
the equilibrium
box overlap distribution function, where the size of  
the box to be used is imposed by the dynamical coherence length.

We have argued that these relations reflect the simpler notions
of the local equilibration in space and the dynamical heterogeneity
of the system, two ingredients which are inherent to 
systems with slow dynamics, independently of their space
dimensionality, their dynamics, or the presence of disorder.

We have emphasized 
throughout the paper that the relations 
(\ref{conj}) - (\ref{conjdriven}) were the physically relevant ones,
as compared to the asymptotic form of Eq.~(\ref{FDRPQ}). Hence, 
they have to be taken into account before drawing conclusions
on the equilibrium glass phase one seeks to study~\cite{fede2}.
This remark, already made in Refs.~\cite{behose,babe}, 
is repeated here because there recently appeared
FD plots built from experimental dynamic measurements in spin 
glasses~\cite{manip3}. As anticipated~\cite{babe}, this dynamical
measurements are very similar to the ones observed in
simulations, despite the enormous difference in time scales 
between experiments and simulations. This is because 
the growth of the coherence length is so slow in spin glasses that
this difference in time scales reduces to a very modest
factor in terms of the coherence length~\cite{zouches,BB,surfing}. 
This in fact implies that the `finite size effects' reported 
in simulations, are present in experiments as well. This points
towards the experimental irrelevance of the thermodynamic limit 
in spin glasses.

Finally, we made clear that the relations 
(\ref{conj}) - (\ref{conjdriven}) had not 
the character of a theorem, meaning
that more work is needed to assess their range of validity.
In particular, coarsening systems are a simple 
counterexample of this relation. This is due to the fact
that the dynamical behaviour is entirely dominated by topological
defects which are absent in the equilibrium simulations, so that
finite time
dynamics and statics do not coincide~\cite{behose,babe,1D}.
There remains to be seen to what extent this alternative `defect' 
description of glassy dynamics generally holds in systems 
like supercooled liquids or soft glassy materials. This is certainly
a quite challenging issue.

\begin{acknowledgments}
M. M\'ezard suggested to use the existing numerical 
results for the box-overlap distributions to weaken the criticism
he had himself formulated~\cite{mezard}. 
I thank E. Marinari, J. Ruiz-Lorenzo,
H. Katzgraber, P. Young who kindly provided me 
with their static data~\cite{boxdata,pqdata3,pqdata4}, 
and A. Barrat, P. Holdsworth,
and M. Sellitto for early discussions and collaborations~\cite{behose,babe}.
J. Kurchan made important remarks 
on a previous version of the manuscript and I warmly thank 
him for this luxury {\it service apr\`es-th\`ese}. This work is supported 
by a European Marie Curie Fellowship No HPMF-CT-2002-01927, CNRS (France)
and Worcester College Oxford. Numerical simulations were
started at the PSMN in ENS Lyon (France) and finished
on Oswell at the Oxford Supercomputing Center, Oxford University. 
\end{acknowledgments}

\end{document}